\documentclass[reprint,superscriptaddress,aps]{revtex4-1}
\usepackage{amsmath}
\usepackage{amssymb}
\usepackage{bm}
\usepackage{braket}
\usepackage{color}
\usepackage[varg]{txfonts}
\usepackage{varwidth}
\usepackage{dcolumn}
\usepackage[breaklinks,colorlinks=true,linkcolor=blue,urlcolor=cyan,citecolor=blue]{hyperref}
\usepackage{graphicx}

\begin{document}

\title{Unified Description of Spin-Lattice Coupling and Thermodynamics in the Pyrochlore Heisenberg Antiferromagnet}

\author{Masaki Gen}
\email{gen@issp.u-tokyo.ac.jp}
\affiliation{Institute for Solid State Physics, The University of Tokyo, Kashiwa 277-8581, Japan}

\author{Hidemaro Suwa}
\email{suwamaro@phys.s.u-tokyo.ac.jp}
\affiliation{Department of Physics, The University of Tokyo, Tokyo 113-0033, Japan}

\author{Shusaku Imajo}
\affiliation{Institute for Solid State Physics, The University of Tokyo, Kashiwa 277-8581, Japan}

\author{Chao Dong}
\affiliation{Institute for Solid State Physics, The University of Tokyo, Kashiwa 277-8581, Japan}

\author{Hiroaki Ueda}
\affiliation{Department of Chemistry, Graduate School of Science, Kyoto University, Kyoto 606-8502, Japan}

\author{Makoto Tachibana}
\affiliation{Research Center for Materials Nanoarchitectonics, National Institute for Materials Science, Tsukuba 305-0044, Japan}

\author{Akihiko Ikeda}
\affiliation{Department of Engineering Science, University of Electro-Communications, Chofu, Tokyo 182-8585, Japan}

\author{Koichi Kindo}
\affiliation{Institute for Solid State Physics, The University of Tokyo, Kashiwa 277-8581, Japan}

\author{Yoshimitsu Kohama}
\affiliation{Institute for Solid State Physics, The University of Tokyo, Kashiwa 277-8581, Japan}

\begin{abstract}

We study an extended model to describe the spin-lattice coupling, incorporating individual vibrations of bonds and atomic sites alongside distance-dependent exchange interactions.
The proposed spin Hamiltonian can be effectively considered as an interpolation between two well-established minimum models, the bond-phonon model and the site-phonon model.
The extended model, which treats bond phonons and site phonons on comparable footing, well reproduces successive field-induced phase transitions as well as the thermodynamic properties of a three-up--one-down state in the pyrochlore-lattice Heisenberg antiferromagnet, including negative thermal expansion, an enhanced magnetocaloric effect, and a sharp specific-heat peak.
The present approach is broadly applicable to various spin models, providing a framework for identifying the primary phonon modes responsible for spin-lattice coupling and for understanding complex magnetic phase diagrams.

\end{abstract}

\date{\today}
\maketitle

{\it Introduction}---The interplay between spin and lattice degrees of freedom, namely the spin-lattice coupling (SLC), is ubiquitous in magnetic compounds due to the variation of exchange parameters against the relevant atomic displacements.
A direct manifestation of the SLC effect is the exchange striction, as represented by the observation of linear magnetostriction proportional to the short-range spin-spin correlation $\langle {\mathbf S}_{i} \cdot {\mathbf S}_{j} \rangle$ on a one-dimensional spin chain \cite{2008_Zap}.
Furthermore, the SLC can be a source of structural instability even with no orbital degrees of freedom, potentially leading to a spin-Peierls(-like) transition characterized by multimer formation \cite{1975_Bra, 1993_Has, 2020_Pau} and/or a spin Jahn-Teller transition accompanied by global lattice deformation \cite{2000_Lee, 2000_Yam, 2002_Tch}.

\begin{figure}[b]
\centering
\includegraphics[width=0.95\linewidth]{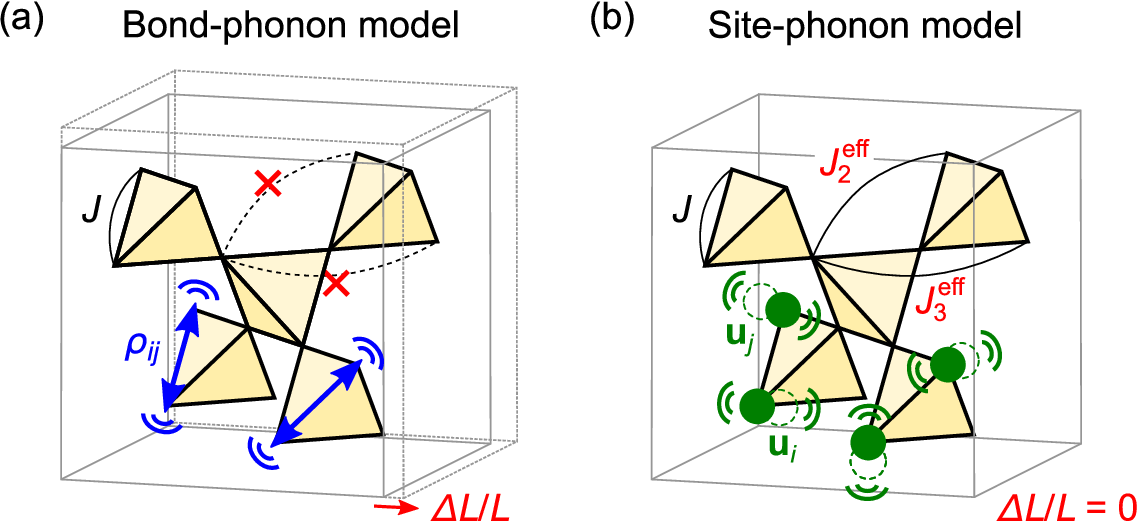}
\caption{Two minimal models describing the SLC: (a) the bond-phonon model assuming independent bond-length change $\rho_{ij}$, and (b) the site-phonon model assuming independent site displacement ${\mathbf u}_{i}$. The BP model does not effectively produce any further-neighbor interactions, while the SP model does not take into account magnetostriction.}
\label{Fig1}
\end{figure}

Two minimal models have been proposed to describe the SLC microscopically: the bond-phonon (BP) model \cite{2004_Pen} and the site-phonon (SP) model \cite{2006_Ber}.
The former assumes independent vibration of each bond, whereas the latter assumes independent displacement of each atomic site (Fig.~\ref{Fig1}).
The BP model has been demonstrated as a powerful approach to reproduce a metamagnetic transition to a magnetization plateau and the associated magnetostriction in frustrated chromium spinels \cite{2004_Pen, 2010_Sha, 2012_Miy, 2014_Kim, 2019_Ros, 2020_Miy, 2021_Yam, 2024_Duc}.
Nevertheless, it oversimplifies the phonon-mediated spin interactions because no further-neighbor interactions beyond the nearest-neighbor ones are considered.
In contrast, the SP model neglects the global lattice deformation but effectively incorporates phonon-mediated three-body spin interactions, enabling the reproduction of magnetic long-range orders (LROs) \cite{2006_Ber, 2007_Mat, 2010_Mat, 2016_Aoy, 2019_Aoy, 2021_Aoy}, complicated phase transitions \cite{2020_Gen, 2023_Gen_1, 2023_Gen_2, 2008_Wan}, and molecular spin excitations \cite{2024_Gao, 2013_Tom, 2018_Gao, 2021_He}.
Recently, the SP model has been applied to a variety of crystallographic systems, such as the triangular lattice \cite{2008_Wan}, kagome lattice \cite{2008_Wan, 2022_Gen}, breathing pyrochlore lattice \cite{2019_Aoy, 2021_Aoy, 2020_Gen, 2023_Gen_1}, and honeycomb lattice \cite{2024_Zhe}, sparking theoretical predictions for the emergence of diverse magnetic phases.

As described above, both the BP and SP models have their own advantages and limitations, thus complementing each other.
Researchers have so far chosen one of these models arbitrarily to explain experimental data, highlighting the need for a more comprehensive description of the SLC for a robust discussion.
As exemplified by the first-principles calculations on the honeycomb-lattice system \cite{2024_Zhe}, the predominant optical phonon modes in real compounds should be renormalized to either BP or SP terms.
Here, we present an extended SLC model by introducing a phenomenological parameter that characterizes a ratio of the SP to the BP contribution, and investigate the magnetic field and temperature dependence of various thermodynamic quantities of the pyrochlore-lattice Heisenberg antiferromagnet.
For comparison, we perform magnetization, magnetostriction, magnetocaloric effect (MCE), and specific heat measurements on a model compound CdCr$_{2}$O$_{4}$ in pulsed high magnetic fields.
We show that many important magnetic and thermodynamic properties can be simultaneously reproduced by a single parameter set, ensuring the validity of the extended SLC model.

{\it Theoretical model}---First, we consider an effective spin Hamiltonian that incorporates both the BP and SP contributions.
While we specifically focus on the classical Heisenberg model on a pyrochlore lattice in this Letter, the present formalization can be applied to any lattice system.
Let $\rho_{ij}$ be the BP displacement between the adjacent sites $i$ and $j$, and let ${\mathbf u}_{i}$ be the SP displacement vector for each site $i$ (Fig.~\ref{Fig1}).
We set antiferromagnetic $J \equiv J(|{\mathbf r}_{ij}^{0}|) > 0$ as the bare nearest-neighbor exchange coupling, where ${\mathbf r}_{ij}^{0} \equiv {\mathbf r}_{i}^{0} - {\mathbf r}_{j}^{0}$ is an equilibrium bond-length without the SLC.
Bearing in mind that the direct exchange interaction dominates in CdCr$_{2}$O$_{4}$, we assume that $J$ depends on the distance between the sites $i$ and $j$.
Using two types of phonon modes, the Hamiltonian of the spin-lattice system is given by
\begin{equation}
\begin{split}
\label{Eq1}
\vspace{-0.2cm}
{\mathcal{H}} =& \sum_{\langle ij\rangle}\left(J(|{\mathbf r}_{i} - {\mathbf r}_{j}|)~{\mathbf S}_{i} \cdot {\mathbf S}_{j} + \frac{k_{\rm BP}}{2}\rho_{ij}^{2}\right) + \sum_{i}\frac{k_{\rm SP}}{2}|{\mathbf u}_{i}|^{2} \\
&- h \sum_{i}S_{i}^{z},
\end{split}
\end{equation}
\vspace{-0.5cm}
\begin{align}
J(|{\mathbf r}_{i} - {\mathbf r}_{j}|) &= J(|{\mathbf r}_{ij}^{0} + \rho_{ij}{\mathbf e}_{ij} + {\mathbf u}_{i} - {\mathbf u}_{j}|) \label{Eq2}\\
&\approx J - J\gamma[\rho_{ij} + {\mathbf e}_{ij} \cdot ({\mathbf u}_{i} - {\mathbf u}_{j})], \label{Eq3}
\vspace{-0.2cm}
\end{align}
where ${\langle ij\rangle}$ runs over all the nearest-neighbor bonds, ${\mathbf S}_{i}$ is a three-dimensional vector spin with unit length at site $i$, $k_{\rm BP} > 0$ ($k_{\rm SP} > 0$) is the spring constant for bond (site) phonons, $h$ is the magnetic field, ${\mathbf e}_{ij} \equiv {\mathbf r}_{ij}^{0}/|{\mathbf r}_{ij}^{0}|$, and $\gamma \equiv -(1/J)(dJ/dr)|_{r=|{\mathbf r}_{ij}^{0}|}$.
For the SP coupling, we take into account a displacement component projected to the bond direction ${\mathbf e}_{ij}$ as the lowest-order contribution to $J$.
Given that the lattice displacement is small enough compared to the lattice constant, we approximate the dependence of $J$ up to the first order of the displacement in Eq.~(\ref{Eq3}).
As the lattice degrees of freedom are quadratic in the Hamiltonian, Eq.~(\ref{Eq1}) is rewritten as
\begin{equation}
\vspace{-0.2cm}
\begin{split}
\label{Eq4}
{\mathcal{H}} &= J\sum_{\langle ij\rangle}{\mathbf S}_{i} \cdot {\mathbf S}_{j} + \frac{k_{\rm BP}}{2}\sum_{\langle ij\rangle}\left((\rho_{ij} - {\overline \rho}_{ij})^{2} - {\overline \rho}_{ij}^{2} \right)\\ &\hspace{+0.3cm}+ \frac{k_{\rm SP}}{2}\sum_{i}\left(({\mathbf u}_{i} - {\overline {\mathbf u}}_{i})^{2} - {\overline {\mathbf u}}_{i}^{2} \right) - h \sum_{i}S_{i}^{z},
\end{split}
\vspace{-0.1cm}
\end{equation}
where ${\overline \rho}_{ij} = (J\gamma/k_{\rm BP}){\mathbf S}_{i} \cdot {\mathbf S}_{j}$ and 
${\overline {\mathbf u}}_{i} = (J\gamma/k_{\rm SP})\sum_{j \in N(i)}{\mathbf e}_{ij}({\mathbf S}_{i} \cdot {\mathbf S}_{j})$, with $N(i)$ denoting the set of neighboring sites of $i$.
By tracing out the BPs and SPs from the Boltzmann distribution through the Gaussian integral, Eq.~(\ref{Eq4}) is exactly reduced to an effective spin Hamiltonian
\begin{equation}
\vspace{-0.2cm}
\begin{split}
\label{Eq5}
{\mathcal{H}}_{\rm eff} &= J\sum_{\langle ij\rangle}\left[{\mathbf S}_{i} \cdot {\mathbf S}_{j} - b({\mathbf S}_{i} \cdot {\mathbf S}_{j})^{2} \right]\\ &\hspace{+0.3cm}- \frac{Jb'}{2}\sum_{i}\sum_{j \neq k \in N(i)}{\mathbf e}_{ij} \cdot {\mathbf e}_{ik}({\mathbf S}_{i} \cdot {\mathbf S}_{j})({\mathbf S}_{i} \cdot {\mathbf S}_{k}) - h \sum_{i}S_{i}^{z},
\end{split}
\vspace{-0.2cm}
\end{equation}
where $b = J\gamma^{2}\left(1/2k_{\rm BP} +  1/k_{\rm SP}\right)$ and $b' = J\gamma^{2}/k_{\rm SP}$, which are the primary control parameters in our model.

Hereafter, we define the ratio of the two SLC parameters as $\eta = b'/b~ (0 \leq \eta \leq 1)$; $\eta = 0$ corresponds to the pure BP model, while $\eta = 1$ corresponds to the pure SP model.
Although the above parametrization of the SLC was proposed in Ref.~\cite{2006_Ber}, the magnetic phase diagram and thermodynamic properties of the extended SLC model [Eq.~(\ref{Eq5})] have yet to be investigated.
To study the thermodynamic properties of Eq.~(\ref{Eq5}), we performed Monte Carlo (MC) simulations for a system size of $N = 16 \times L^{3}$ sites with $L = 4$ under periodic boundary conditions.
For simplicity, the local spin length was fixed to unity, $|{\mathbf S}_{i}| = 1$.
Further details of the simulation methods are provided in the Supplemental Material \cite{Supple}.

{\it Magnetic properties of CdCr$_{2}$O$_{4}$}---Chromium spinel oxides {\it A}Cr$_{2}$O$_{4}$ ({\it A} = Mg, Zn, Cd, Hg), where Cr$^{3+}$ ions with $S = 3/2$ form a pyrochlore lattice, have been the focus of much attention for studying the effect of SLC in geometrically frustrated magnets \cite{2012_Miy, 2014_Kim, 2019_Ros, 2007_Mat, 2010_Mat, 2016_Aoy, 2019_Aoy, 2021_Aoy, 2024_Gao, 2013_Tom, 2018_Gao, 2021_He, 2005_Chu, 2009_Ji, 2005_Ued, 2006_Ued, 2011_Miy, 2013_Miy, 2014_Miy, 2014_Nak, 2015_Kim, 2022_Kim, 2013_Kit, 2006_Fen, 2011_Xia, 2012_Kum, 2016_Wys}.
As the orbital degree of freedom is quenched due to the $3d^{3}$ electronic configuration under the octahedral crystal field, the SLC acts as a major perturbation to resolve the magnetic frustration, leading to magnetostructural transitions at low temperatures \cite{2007_Mat, 2018_Gao, 2005_Chu, 2009_Ji} and in high magnetic fields \cite{2012_Miy, 2014_Kim, 2019_Ros, 2005_Ued, 2006_Ued, 2011_Miy, 2013_Miy, 2014_Miy, 2014_Nak, 2015_Kim, 2022_Kim}.

\begin{figure}[t]
\centering
\includegraphics[width=\linewidth]{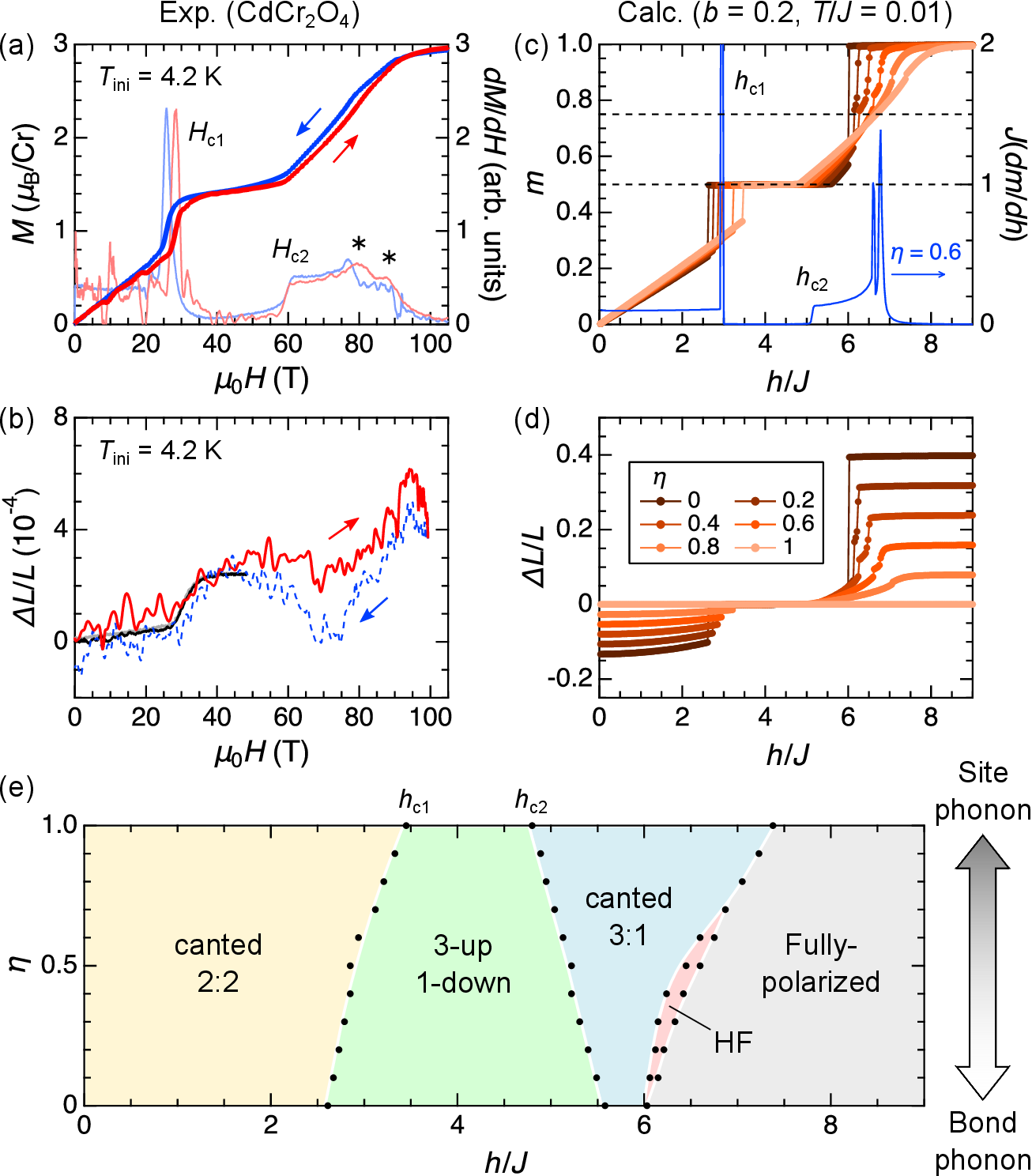}
\caption{(a),(b) Experimental (a) magnetization and (b) magnetostriction curves of polycrystalline CdCr$_{2}$O$_{4}$ at $T_{\rm ini} = 4.2$~K obtained using the single-turn-coil technique, where the sample temperature is expected to be under (quasi)adiabatic conditions. In panel (a), $dM/dH$ is displayed by thin colors in the right axis. In panel (b), the data up to 50~T (black) are obtained in a nondestructive pulsed magnet. (c),(d) Calculated (c) magnetization and (d) magnetostriction curves for the extended SLC model [Eq.~(\ref{Eq5})] with $b = 0.2$ and various values of $\eta$ at $T/J = 0.01$. In (a) and (c), the field derivative of the magnetization (for $\eta = 0.6$) is plotted in the right axis. (e) Theoretical phase diagram of Eq.~(\ref{Eq5}) with $b = 0.2$ as a function of magnetic field $h$ and the SP contribution $\eta$ at $T/J = 0.01$.}
\label{Fig2}
\end{figure}

CdCr$_{2}$O$_{4}$ undergoes a magnetic transition at $T_{\rm N} \approx 7.6$~K in zero field \cite{2013_Kit}, accompanied by a cubic-to-tetragonal structural transition \cite{2005_Chu}. 
Figure~\ref{Fig2}(a) shows a magnetization process of CdCr$_{2}$O$_{4}$ at the initial temperature of $T_{\rm ini} = 4.2$~K, where a 1/2-magnetization plateau appears between $\mu_{0}H_{\rm c1} = 28$~T and $\mu_{0}H_{\rm c2} = 58$~T, accompanied by a sharp metamagnetic transition.
A high-field neutron-diffraction study revealed the emergence of a three-up--one-down LRO above $H_{\rm c1}$ \cite{2010_Mat}, which should be stabilized by the SLC favoring a collinear spin configuration via the biquadratic interaction $-({\mathbf S}_{i} \cdot {\mathbf S}_{j})^{2}$.
With further increasing of a magnetic field above $H_{\rm c2}$, a double-peak anomaly appears in $dM/dH$ around 80--88 T, as indicated by asterisks in Fig.~\ref{Fig2}(a), and then spins are fully polarized at $\mu_{0}H_{\rm sat} \approx 90$~T.
The presence of an additional high-field (HF) phase immediately below saturation was also indicated by a previous magneto-optical spectroscopy of the $d$--$d$ transitions and the exciton-magnon-phonon transitions \cite{2013_Miy}.
The above-mentioned successive phase transitions are universally observed for other {\it A}Cr$_{2}$O$_{4}$ families  \cite{2005_Ued, 2006_Ued, 2011_Miy, 2013_Miy, 2014_Miy}.

We consider CdCr$_{2}$O$_{4}$ as a suitable compound for verifying the extended SLC model because of the relatively low saturation field and the availability of high-quality single crystals.
To comprehend the thermodynamic properties of CdCr$_{2}$O$_{4}$, we measured magnetostriction, MCE, and specific heat in pulsed high magnetic fields by utilizing recently developed experimental techniques \cite{2017_Ike, 2013_Kih, 2015_Koh, 2021_Mat, 2021_Ima, 2022_Koh} at ISSP, University of Tokyo.
Details of the experimental methods are presented in the Supplemental Material \cite{Supple}.

{\it Magnetic phase diagram of the extended SLC model}---Figures~\ref{Fig2}(a)--\ref{Fig2}(d) compare the experimental magnetization and magnetostriction curves of CdCr$_{2}$O$_{4}$ with simulations, where we use different $\eta$ values with $b = 0.2$ (for results with other values of $b$, see Fig.~S3 \cite{Supple}).
The corresponding theoretical phase diagram as a function of magnetic field and $\eta$ is shown in Fig.~\ref{Fig2}(e).
Throughout the Letter, $\Delta L/L$ denotes the relative change in sample length, referred to as magnetostriction (thermal expansion) when plotted as a function of field (temperature).
In the theoretical model, the saturation field is $h/J = 8$ without the SLC.
For both the BP and SP models, a first-order transition from a canted 2:2 to a three-up--one-down state occurs at a lower critical field of the 1/2-magnetization plateau ($h_{\rm c1}$), while a second-order transition from the three-up--one-down to a canted 3:1 state occurs at an upper critical field ($h_{\rm c2}$).
These trends agree with the experimental magnetization curve [Fig.~\ref{Fig2}(a)] \cite{comment1}.
Furthermore, the theoretically predicted positive magnetostriction (except for the case of $\eta = 1$), where the lattice expansion is rather enhanced above $h_{\rm c2}$, also agrees with the experimental observation [Fig.~\ref{Fig2}(b)] (see also Fig.~S3 for the magnetostriction curve of HgCr$_{2}$O$_{4}$ \cite{Supple}).
In the BP model ($\eta = 0$), the plateau width increases with increasing $b$ \cite{2004_Pen}.
However, the plateau width decreases with increasing $\eta$; for $b = 0.2$, $(h_{\rm c2}-h_{\rm c1})/8J \approx 0.31$ for $\eta = 0$ and 0.17 for $\eta = 1$ [Fig.~\ref{Fig2}(c)].

Neither the conventional BP nor SP models can account for the double-peak anomaly in $dM/dH$ observed in CdCr$_{2}$O$_{4}$ [Fig.~\ref{Fig2}(a)], and the emergence of a spin-nematic phase driven by quantum effects has been proposed \cite{2011_Miy, 2014_Nak, 2015_Tak}.
Remarkably, this high-field feature in CdCr$_{2}$O$_{4}$ can be explained by our extended SLC model, even under the assumption of classical spins.
As shown in Fig.~\ref{Fig2}(e), an additional HF phase emerges immediately above the canted 3:1 phase for $0 < \eta < 0.7$.
The field derivative of magnetization for $\eta = 0.6$ [blue curve in Fig.~\ref{Fig2}(c)] successfully reproduces the double-peak structure in CdCr$_{2}$O$_{4}$.
Note that, in our calculations, the two-step magnetization jump is rapidly smeared out and eventually disappears with increasing temperature (see Fig.~S4 \cite{Supple}).
In the HF phase between the two $dM/dH$ peaks, we identify the emergence of a magnetic LRO state characterized by a periodic stacking along the $\langle 111 \rangle$ axis, consisting of a triangular layer with a 120$^{\circ}$ spin configuration, an all-up kagome layer, all-up triangular layer, and another all-up kagome layer (see Fig.~S6 for the detailed magnetic structure and lattice distortion \cite{Supple}).
The existence of a phase transition into the HF phase is further supported by temperature-dependent simulations at fixed fields in the range $h/J = 6.6$--$6.75$, where the transition temperature is found to be $T_{\rm N}/J \approx 0.01$ at $h/J = 6.75$ (see Fig.~S5 \cite{Supple}).

\begin{figure}[t]
\centering
\includegraphics[width=\linewidth]{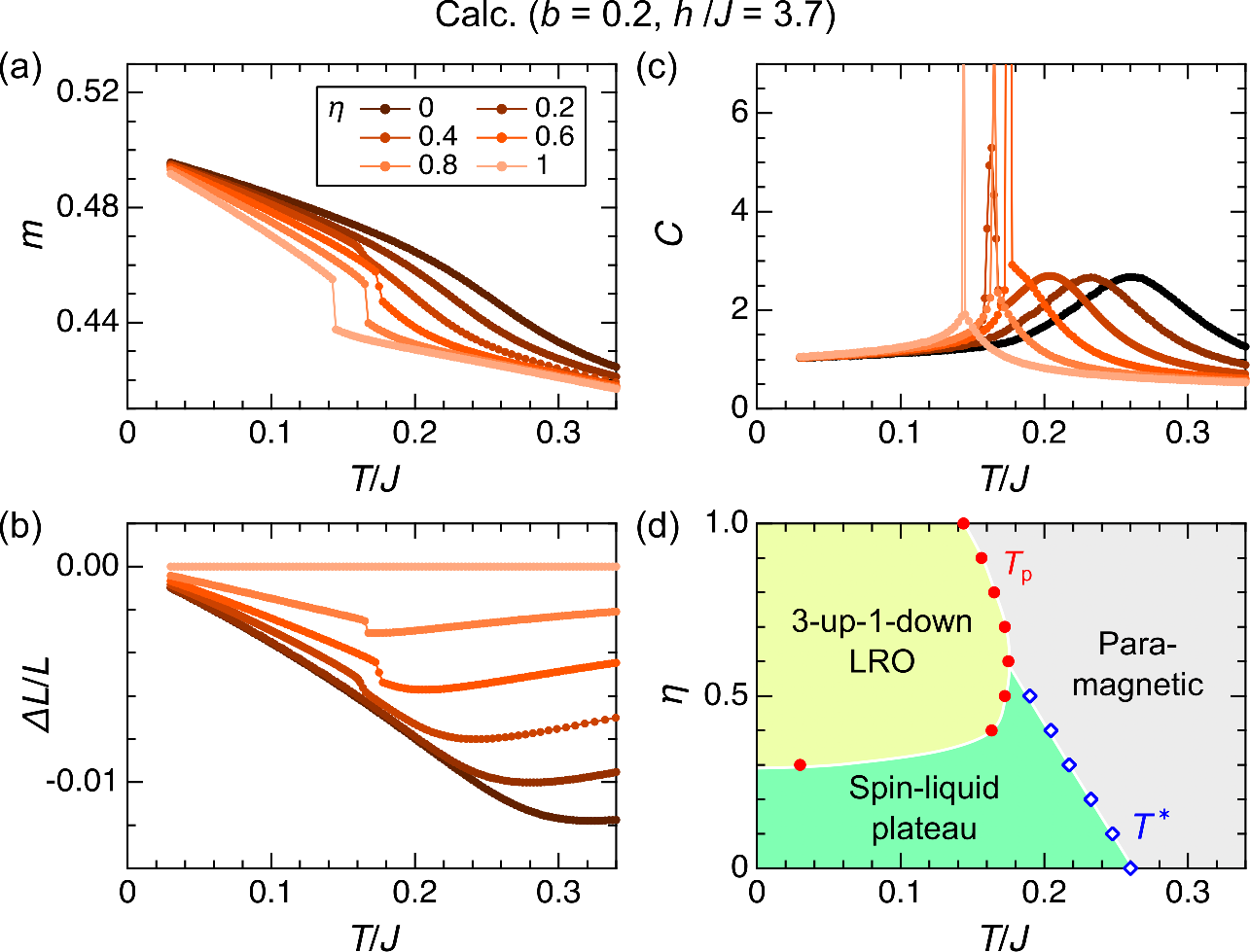}
\caption{[(a)--(c)] Temperature dependence of (a) magnetization, (b) thermal expansion, and (c) specific heat for the extended SLC model [Eq.~(\ref{Eq5})] with $b = 0.2$ and various $\eta$ values in the low-field side of the 1/2-magnetization plateau at $h/J = 3.7$. (d) Theoretical phase diagram of Eq.~(\ref{Eq5}) with $b = 0.2$ as a function of temperature $T$ and the site-phonon contribution $\eta$ at $h/J = 3.7$. A phase transition to a three-up--one-down LRO state (crossover to a spin-liquid plateau state) is characterized by a sharp (broad) specific-heat peak, which is indicated by red circles (open blue diamonds).}
\label{Fig3}
\end{figure}

{\it Thermodynamic properties in the 1/2-plateau phase}---Figures~\ref{Fig3}(a)--\ref{Fig3}(c) show the calculated temperature dependence of magnetization, thermal expansion, and specific heat for various $\eta$ values with $b = 0.2$ at $h/J = 3.7$.
Figure~\ref{Fig3}(d) shows the corresponding phase diagram as a function of temperature and $\eta$.
In the BP limit ($\eta = 0$), the specific heat exhibits a broad peak at $T^{*}/J \approx 0.26$, indicating a crossover from the paramagnetic state to the spin-liquid plateau state, as reported in Refs.~\cite{2010_Sha, 2021_Aoy}.
Below the crossover temperature $T^{*}$, each tetrahedron adopts a three-up--one-down configuration with local $T_{2}$ symmetry, while macroscopic degeneracy persists across the entire system.
As $\eta$ increases, $T^{*}$ gradually decreases, and another sharp specific-heat peak emerges at a lower temperature $T_{\rm p}$ for $\eta \approx 0.3$, signaling a first-order transition to a three-up--one-down LRO state.
The transition temperature $T_{\rm p}$ increases with increasing $\eta$ and eventually converges with $T^{*}$ for $\eta \approx 0.6$.
Then, $T_{\rm p}$ gradually decreases as $\eta$ further increases, resulting in $T_{\rm p}/J \approx 0.14$ in the SP limit ($\eta = 1$), consistent with Ref.~\cite{2021_Aoy}.

\begin{figure}[t]
\centering
\includegraphics[width=\linewidth]{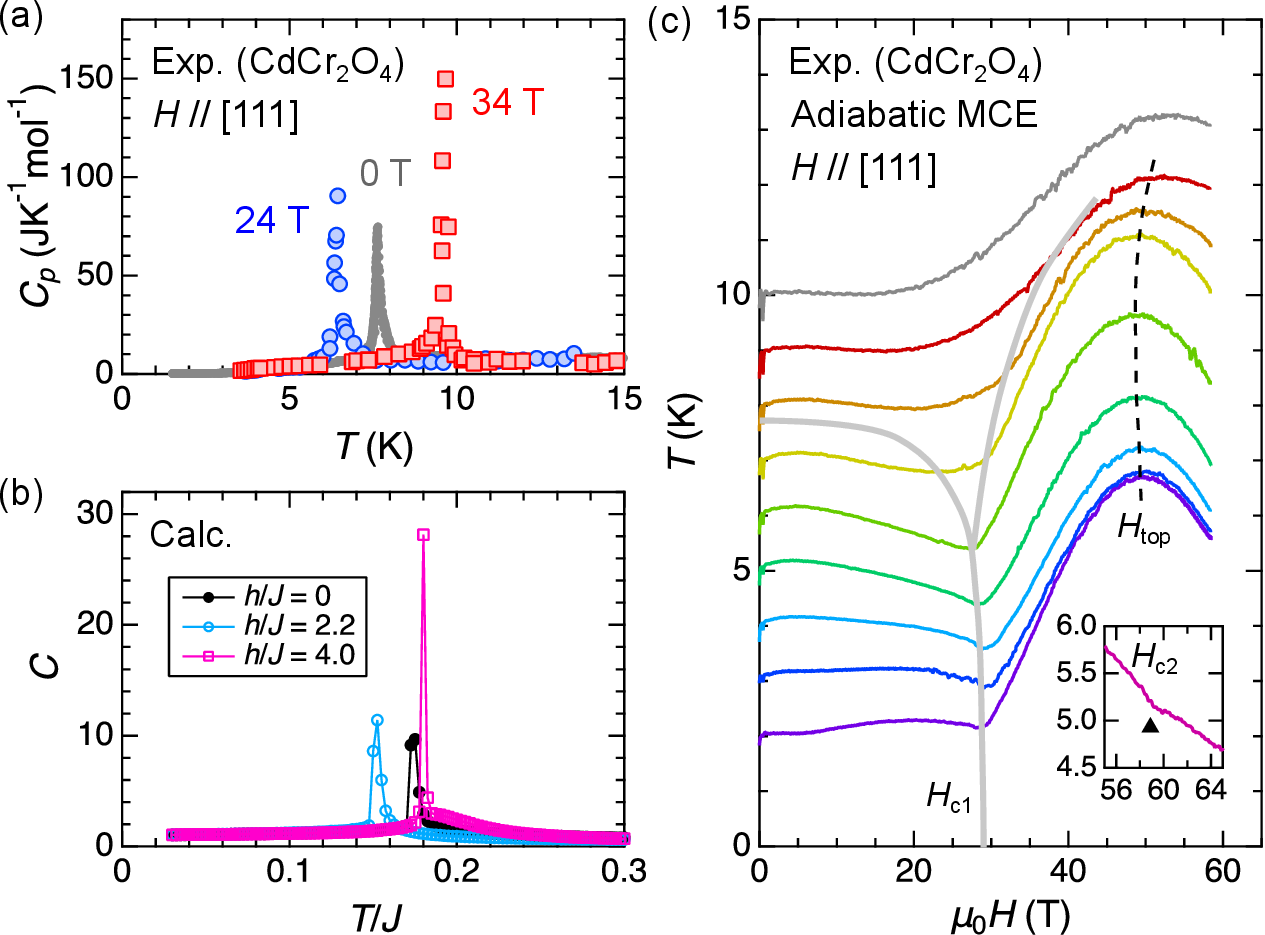}
\caption{(a) Temperature dependence of specific heat at 24~T and 34~T for $H \parallel [111]$ in CdCr$_{2}$O$_{4}$, obtained using a flat-top nondestructive long pulsed magnet \cite{2015_Koh, 2021_Mat, 2021_Ima, 2022_Koh}. (b) Temperature dependence of specific heat at several $h$ values for the extended SLC model [Eq.~(\ref{Eq5})] with $b = 0.2$ and $\eta = 0.6$. (c) Magnetocaloric effect (MCE) for $H \parallel [111]$ in CdCr$_{2}$O$_{4}$, measured under adiabatic conditions in a nondestructive pulsed magnet. The inset shows a magnified view of the $T(H)$ curve around $H_{\rm c2}$ measured using a different setup. All the data correspond to the field-increasing process. The thick gray lines indicate the phase boundaries separating the canted 2:2, three-up--one-down, and paramagnetic phases.}
\label{Fig4}
\end{figure}

We now discuss the behavior of the specific heat in applied magnetic fields.
Figure~\ref{Fig4}(a) shows the specific heat data of CdCr$_{2}$O$_{4}$ at 24~T and 34~T for $H \parallel [111]$ (see Fig.~S7 for the data obtained from polycrystalline samples \cite{Supple}).
At 24~T ($< H_{\rm c1}$), the transition temperature shifts to a lower value compared to $T_{\rm N} = 7.6$~K at zero field.
Remarkably, a more pronounced peak is observed at $T_{\rm p} = 9.7$~K at 34~T ($> H_{\rm c1}$), indicating the appearance of three-up--one-down LRO below $T_{\rm p}$.
A similar behavior has also been reported for polycrystalline HgCr$_{2}$O$_{4}$ \cite{2022_Kim}.
In the SP model, the three-body quadratic terms of the form $-({\mathbf S}_{i} \cdot {\mathbf S}_{j})({\mathbf S}_{i} \cdot {\mathbf S}_{k})$ in Eq.~(\ref{Eq5}) effectively produce second- and third-nearest-neighbor antiferromagnetic interactions, $J_{2}^{\rm eff}$ and $J_{3}^{\rm eff}$ [Fig.~\ref{Fig1}(b)].
Under the three-up--one-down constraint, $J_{3}^{\rm eff}$ dominates over $J_{2}^{\rm eff}$, as  $J_{3}^{\rm eff} = 2J_{2}^{\rm eff}$, leading to a three-up--one-down LRO with cubic $P4_{3}32$ symmetry rather than rhombohedral $R{\overline 3}m$ symmetry \cite{2006_Ber}.
This is consistent with experimental observations in HgCr$_{2}$O$_{4}$ and CdCr$_{2}$O$_{4}$ based on high-field neutron diffraction \cite{2007_Mat, 2010_Mat}.
As mentioned above, our calculations reveal that, in the extended SLC model, $\eta$ must be at least 0.3 to induce the three-up--one-down LRO.
Moreover, our specific heat data at 34~T does not exhibit a hump structure above $T_{\rm p}$, which is consistent with the calculated one for $\eta \gtrapprox 0.6$ [Fig.~\ref{Fig3}(c)].

The transition temperature to the 1/2-plateau phase at 34~T, $T_{\rm p}$, is higher than $T_{\rm N}$ at zero field, suggesting the stabilization of collinear spin configuration by thermal fluctuations \cite{2019_Ros}.
This trend cannot be reproduced by the pure SP model ($\eta = 1$), as shown in Fig.~6 in Ref.~\cite{2021_Aoy}.
This discrepancy can be resolved by considering the extended model that incorporates the additional BP contributions.
Figure~\ref{Fig4}(b) shows the calculated specific heat at several $h$ values for Eq.~\eqref{Eq5} with $b = 0.2$ and $\eta = 0.6$.
The transition temperature to the three-up--one-down LRO phase at $h/J=4.0$ is higher than that to the (canted) 2:2 LRO phase at $h/J = 0$ and 2.2.
The corresponding specific-heat peak at $h/J=4.0$ is also the sharpest, supporting the validity of the choice $\eta = 0.6$ for reproducing the experimentally observed specific-heat behavior.

Finally, we turn to the negative thermal expansion (NTE) that emerges in the 1/2-plateau phase.
A previous thermal expansion measurement in CdCr$_{2}$O$_{4}$ reported the NTE on the low-field side of the 1/2-plateau phase at 30~T, accompanied by a discontinuous jump in thermal expansion at the transition to the paramagnetic phase \cite{2019_Ros}.
These behaviors are well reproduced by our extended SLC model for intermediate $\eta$ values with $b = 0.2$ at $h/J = 3.7$ (close to $h_{\rm c1}$) [Fig.~\ref{Fig3}(b)].
In Ref.~\cite{2019_Ros}, the NTE is attributed to a negative change in magnetization with temperature, i.e., ${\partial m}/{\partial T} < 0$, arising from a nearly localized band of spin excitations in the 1/2-plateau phase.
Our MC simulations indicate that the temperature range exhibiting NTE approximately corresponds to the region where either the three-up--one-down LRO phase or the spin-liquid plateau phase emerges.
We also theoretically predict that, on the high-field side of the 1/2-plateau phase, the system exhibits positive thermal expansion down to the lowest temperature due to the spin-gap closing on approaching $h_{\rm c2}$ (see Fig.~S8 \cite{Supple}), although this behavior has not yet been observed experimentally.

From a thermodynamic point of view, the NTE behavior is closely related to an enhanced MCE, as proposed in Ref.~\cite{2019_Ros}.
According to thermodynamic relations, the field derivative of the sample temperature $T(H)$ under adiabatic conditions is expressed as
\begin{equation}
\label{Eq6}
\vspace{-0.1cm}
\bigg( \frac{\partial T}{\partial H} \bigg)_{S} = -\frac{T}{C_{p}}\bigg( \frac{\partial M}{\partial T} \bigg)_{H}.
\end{equation}
Figure~\ref{Fig4}(c) shows the adiabatic MCE data of CdCr$_{2}$O$_{4}$ during the field-up sweep for $H \parallel [111]$, measured at various initial temperatures.
Below $T_{\rm N}$, the $T(H)$ curve exhibits a dip at $H_{\rm c1}$, indicating the increase in magnetic entropy.
Upon entering the 1/2-plateau phase, the $T(H)$ curves develop a domelike structure with a maximum near $\mu_{0}H_{\rm top} \approx 50$~T, where $T(H)$ increases by up to 5~K.
The upper phase boundary at $H_{\rm c2}$ is also visible as a kink in the $T(H)$ curve [inset of Fig.~\ref{Fig4}(c)], beyond which sample cooling persists in the higher field region, at least up to 65~T.
These MCE data suggest that the sign change in ${\partial M}/{\partial T}$ from negative to positive occurs at a magnetic field higher than the midpoint of the 1/2-plateau phase, i.e., $\mu_{0}H_{\rm top} > \mu_{0}(H_{\rm c1} + H_{\rm c2})/2 \approx 43$~T \cite{comment2}.
This trend is consistent with the experimental magnetization curves \cite{2005_Ued}, and is also reproduced by our extended SLC model (see Fig.~S4 in the SM \cite{Supple}).
The comparison with these MCE results demonstrates that our extended SLC model consistently captures all the thermodynamic properties in the 1/2-plateau phase of CdCr$_{2}$O$_{4}$.

{\it Conclusion}---We have validated the extended SLC model for the pyrochlore-lattice Heisenberg antiferromagnet by demonstrating its consistency with the thermodynamic properties of CdCr$_{2}$O$_{4}$.
The extended SLC model interpolates between the BP and SP models, characterized by a phenomenological parameter $\eta$, which describes the ratio of SP modes.
Our MC simulations reveal that introducing both the BP and SP modes on comparable footing, e.g., $\eta = 0.6$, provides the best agreement with experimental observations, including negative thermal expansion, an enhanced MCE, and a sharp specific heat peak in the 1/2-plateau phase, as well as a two-step phase transition just below the saturation field.
This theoretical framework offers a practical approach for testing the primary phonon modes responsible for SLC without the need for sophisticated techniques such as first-principles calculations.
Furthermore, applying the extended SLC model to other lattice systems with strong SLC may provide new insights into the complex phase diagrams  \cite{2025_Gru}.

{\it Acknowledgments}---Calculations were performed using computational resources from the Supercomputer Center at the Institute for Solid State Physics, the University of Tokyo.
This work was financially supported by the JSPS KAKENHI Grants-In-Aid for Scientific Research (No.~20J10988, No.~24H01609, and No.~24H01633).
The authors thank A. Zampa for his kind support with the specific heat measurements, and K. Penc, N. Shannon, and A. Samanta for fruitful discussions.


\begin{thebibliography}{99}
\bibitem{comment1} Although the experimental $M$--$H$ curves [Fig.~\ref{Fig2}(a)] for up and down sweeps do not perfectly overlap near $H_{\rm c2}$, we attribute this discrepancy to the limited accuracy of the present magnetization measurement in the single-turn-coil system as well as the MCE, rather than to an intrinsic first-order character.
\end{thebibliography}
\end{document}


\title{Supplemental Material for\\``Unified Description of Spin-Lattice Coupling and Thermodynamics in the Pyrochlore Heisenberg Antiferromagnet''}

\author{Masaki Gen}
\email{gen@issp.u-tokyo.ac.jp}
\affiliation{Institute for Solid State Physics, The University of Tokyo, Kashiwa 277-8581, Japan}

\author{Hidemaro Suwa}
\email{suwamaro@phys.s.u-tokyo.ac.jp}
\affiliation{Department of Physics, The University of Tokyo, Tokyo 113-0033, Japan}

\author{Shusaku Imajo}
\affiliation{Institute for Solid State Physics, The University of Tokyo, Kashiwa 277-8581, Japan}

\author{Chao Dong}
\affiliation{Institute for Solid State Physics, The University of Tokyo, Kashiwa 277-8581, Japan}

\author{Hiroaki Ueda}
\affiliation{Department of Chemistry, Graduate School of Science, Kyoto University, Kyoto 606-8502, Japan}

\author{Makoto Tachibana}
\affiliation{Research Center for Materials Nanoarchitectonics, National Institute for Materials Science, Tsukuba 305-0044, Japan}

\author{Akihiko Ikeda}
\affiliation{Department of Engineering Science, University of Electro-Communications, Chofu, Tokyo 182-8585, Japan}

\author{Koichi Kindo}
\affiliation{Institute for Solid State Physics, The University of Tokyo, Kashiwa 277-8581, Japan}

\author{Yoshimitsu Kohama}
\affiliation{Institute for Solid State Physics, The University of Tokyo, Kashiwa 277-8581, Japan}

\maketitle

\onecolumngrid

\vspace{+1.0cm}
This Supplemental Material includes contents as listed below:\\
\\
Note 1.~Monte Carlo simulations\\
Note 2.~Experimental methods\\
Note 3.~Choice of the SLC parameter $b$\\
Note 4.~System-size dependence of calculated magnetization curves\\
Note 5.~Magnetostriction curve of HgCr$_{2}$O$_{4}$ up to 50~T\\
Note 6.~Temperature dependence of calculated magnetization and magnetostriction curves\\
Note 7.~Magnetic structure of the HF phase\\
Note 8.~Specific heat data for polycrystalline CdCr$_{2}$O$_{4}$ in high magnetic fields\\
Note 9.~Thermodynamic properties on the high-field side of the 1/2-plateau phase\\

\newpage
\subsection*{\label{Sec1} \large Note 1. Monte Carlo simulations}

We performed Monte Carlo (MC) simulations for the extended spin-lattice coupling (SLC) model,
\begin{equation}
\label{EqS1}
{\mathcal{H}}_{\rm eff} = J\sum_{\langle ij\rangle}\left[{\mathbf S}_{i} \cdot {\mathbf S}_{j} - b({\mathbf S}_{i} \cdot {\mathbf S}_{j})^{2} \right] - \frac{Jb'}{2}\sum_{i}\sum_{j \neq k \in N(i)}{\mathbf e}_{ij} \cdot {\mathbf e}_{ik}({\mathbf S}_{i} \cdot {\mathbf S}_{j})({\mathbf S}_{i} \cdot {\mathbf S}_{k}) - h \sum_{i}S_{i}^{z},
\end{equation}
which is equivalent to Eq.~(5) in the main text, in order to compute several physical quantities: magnetization $m$, magnetostriction/thermal expansion $\Delta L/L$, and specific heat $C$.
Each physical quantity is calculated by
\renewcommand{\theequation}{S\arabic{equation}}
\begin{align}
    m &= \frac{1}{N}\sum_i \langle \mathbf{S}_i \cdot \mathbf{h} \rangle,\\
    \frac{\Delta L}{L} &\propto \frac{1}{N_b}\sum_{\langle ij \rangle}\langle \rho_{ij} \rangle = \frac{J \gamma}{N_b k_{\rm BP}}\sum_{\langle ij \rangle}\langle  \mathbf{S}_i \cdot \mathbf{S}_j\rangle \label{eq_ms},\\
    C &= \frac{\langle \mathcal{H}_{\rm eff}^2 \rangle - \langle \mathcal{H}_{\rm eff}\rangle^2}{NT^2},
\end{align}
where $\mathbf{h}$ is the magnetic field, and $N$ and $N_b$ are the number of sites and bonds, respectively.
In Eq.~\eqref{eq_ms}, $\Delta L/L$ is proportional to the average displacement of bond phonons $\rho_{ij}$, as verified in several spinel compounds \cite{2020_Miy}.
The absolute values of the lattice displacements, $\rho_{ij}$ and $\mathbf{u}_i$, are arbitrary in our model because their coefficient can be absorbed by the renormalization of the parameters, $\gamma$, $k_{\rm BP}$, and $k_{\rm SP}$, keeping the ratios $\gamma^2/k_{\rm BP}$ and $\gamma^2/k_{\rm SP}$.

Following Ref.~\cite{2022_Gen}, we retained the lattice degrees of freedom in the simulation and introduced microcanonical updates to accelerate the thermalization process and shorten the autocorrelation time.
The random or the over-relaxation-like update step alternately follows five microcanonical update MC steps. 
The single MC step is composed of $N$ local spin and site-phonon updates and $N_b$ local bond-phonon updates.
More than $2^{22}$ MC steps were run for each parameter set, and the latter half was used to calculate the averages of the physical quantities.
We also combined the replica-exchange MC method \cite{1996_Huk}, or parallel tempering, to avoid trapping in local minima at low temperatures.
The error bars of the Monte Carlo simulations are smaller than the symbol sizes in all plots, as estimated by standard binning (blocking) analysis.

\vspace{+2.0cm}
\subsection*{\label{Sec2} \large Note 2. Experimental methods}

Single crystals of CdCr$_{2}$O$_{4}$ were synthesized by a flux method using PbO as the flux \cite{2013_Kit}.
Polycrystalline powder samples of CdCr$_{2}$O$_{4}$ were synthesized by conventional solid-state reactions \cite{2006_Ued}.
Polycrystalline powder samples of HgCr$_{2}$O$_{4}$ were prepared by a thermal decomposition of Hg$_{2}$CrO$_{4}$ in an evacuated silica tube \cite{2006_Ued}.

The longitudinal magnetostriction up to 50~T and 100~T was measured by the optical fiber-Bragg-grating (FBG) method in the nondestructive pulsed magnet (36~ms duration) and the vertical single-turn-coil (STC) system (8~$\mu$s duration), respectively.
A relative sample-length change $\Delta L/L$ was detected by the optical filter method \cite{2017_Ike}.
The magnetocaloric effect (MCE) up to 59~T and 65~T was measured under quasiadiabatic conditions for $H \parallel [111]$ using the nondestructive pulsed magnet with the field duration of 36~ms and 11~ms, respectively.
Here, the quasiadiabatic conditions were achieved by removing helium gas in the sample space.
A sensitive Au$_{16}$Ge$_{84}$ film thermometer was sputtered on the (111) plane of the CdCr$_{2}$O$_{4}$ single crystal \cite{2013_Kih}.
The specific heat $C_{p}$ of CdCr$_{2}$O$_{4}$ up to 34~T was measured using the quasiadiabatic method under a flat-top long pulsed field \cite{2015_Koh, 2021_Ima}.
All of the experiments were performed at the Institute for Solid State Physics, University of Tokyo, Japan.

\newpage
\subsection*{\label{Sec3} \large Note 3. Choice of the SLC parameter $b$}

\begin{figure}[h]
\centering
\vspace{+0.5cm}
\includegraphics[width=0.4\linewidth]{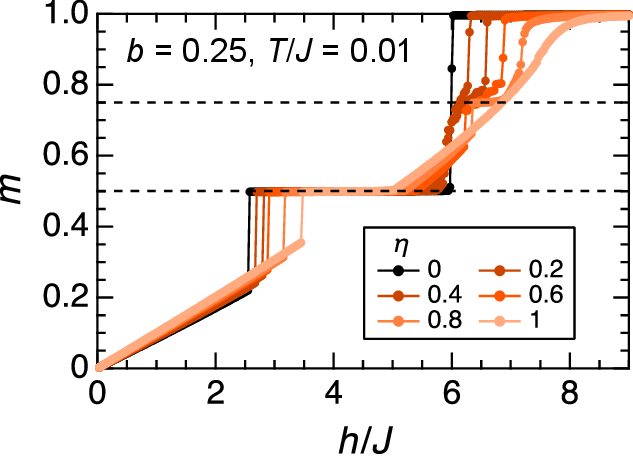}
\caption{Calculated magnetization curves for the extended SLC model [Eq.~\eqref{EqS1}] with $b = 0.25$ and various values of $\eta$ at $T/J = 0.01$.}
\label{FigS1}
\end{figure}

We selected $b = 0.2$ as a typical value of the spin-lattice coupling (SLC) parameter in this study because it provides the best overall agreement with the key experimental observations in CdCr$_{2}$O$_{4}$ and HgCr$_{2}$O$_{4}$.
Bergman {\it et al}. reported calculations at $b = 0.1$ for $\eta = 0$--0.5 \cite{2006_Ber}.
However, the relative field range of the 1/2-magnetization plateau for $b = 0.1$ is rather narrower than observed in CdCr$_{2}$O$_{4}$ and HgCr$_{2}$O$_{4}$.
Conversely, increasing the SLC parameter to $b = 0.25$ improves the width of the 1/2-plateau but overenhances the magnetization anomaly just below the saturation field, as shown in Fig.~\ref{FigS1}.
Balancing these two aspects, we conclude that $b = 0.2$ yields the most faithful overall reproduction of the magnetization curves for CdCr$_{2}$O$_{4}$ and HgCr$_{2}$O$_{4}$.

\vspace{+2.0cm}
\subsection*{\label{Sec4} \large Note 4. System-size dependence of calculated magnetization curves}

\begin{figure}[h]
\centering
\vspace{+0.5cm}
\includegraphics[width=0.4\linewidth]{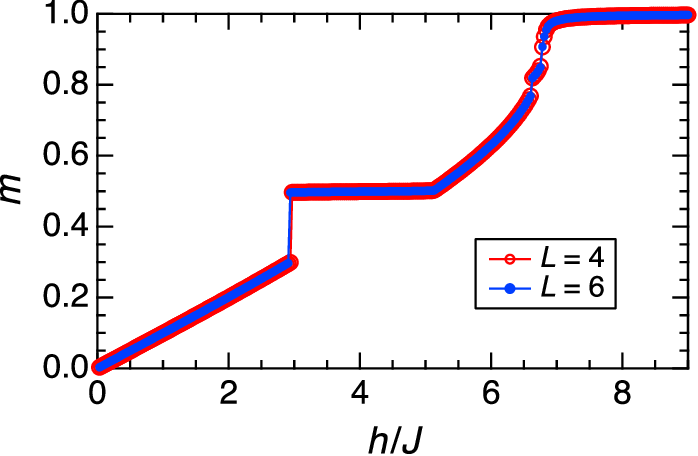}
\caption{Calculated magnetization curves for the extended SLC model [Eq.~\eqref{EqS1}] with $b = 0.2$, $\eta = 0.6$ at $T/J = 0.01$ for $L = 4$ (red) and $L = 6$ (blue).}
\label{FigS2}
\end{figure}

In this study, we mainly performed MC simulations for a system size of $N = 16 \times L^{3}$ with $L = 4$.
We confirmed that the results remain essentially unchanged for $L = 6$.
For example, Fig.~\ref{FigS2} compares the calculated magnetization curves with $b = 0.2$ and $\eta = 0.6$ at $T/J = 0.01$ for $L = 4$ and $L = 6$, showing that both the width of the 1/2-magnetization plateau and the presence of the HF phase are reproduced in both cases.

\newpage
\subsection*{\label{Sec5} \large Note 5. Magnetostriction curve of HgCr$_{2}$O$_{4}$ up to 50~T}

\begin{figure}[h]
\centering
\vspace{+0.5cm}
\includegraphics[width=0.35\linewidth]{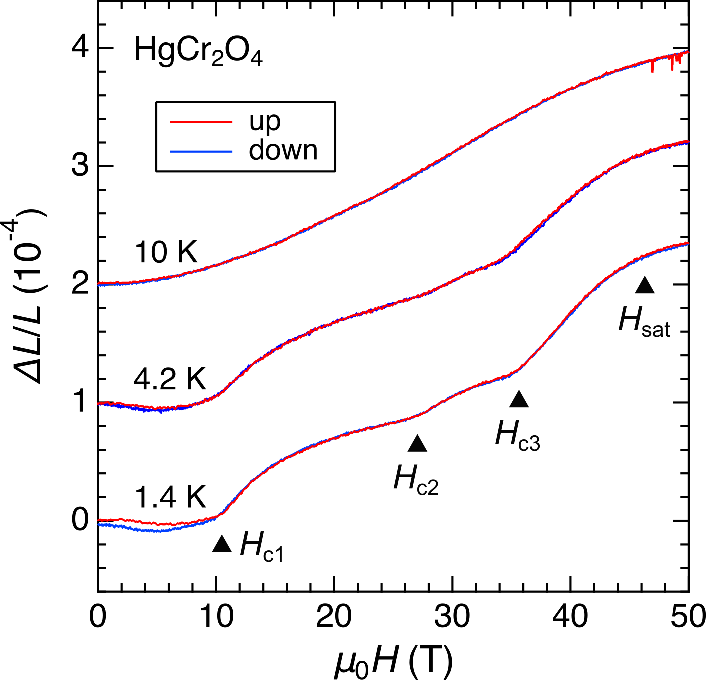}
\caption{Magnetostriction curves of polycrystalline HgCr$_{2}$O$_{4}$ at several temperatures obtained using the nondestructive pulsed magnet.}
\label{FigS3}
\end{figure}

Figure~\ref{FigS3} shows the longitudinal magnetostriction curves of polycrystalline HgCr$_{2}$O$_{4}$ measured below and above $T_{\rm N}=5.8$~K.
At $T = 1.4$ and 4.2~K, a downward convex behavior is observed below $\mu_{0}H_{\rm c1} = 10$~T, where a metamagnetic transition to a 1/2-plateau phase takes place \cite{2006_Ued}.
Subsequently, a dramatic increase in $\Delta L/L$ is observed at each phase transition field up to the saturation field, $\mu_{0}H_{\rm sat} \approx 46$~T, namely at $\mu_{0}H_{\rm c2} = 27$~T, $\mu_{0}H_{\rm c3} = 36$~T.
Notably, the total change in $\Delta L/L$ between $H_{\rm c2}$ and $H_{\rm sat}$ is approximately twice as large as that below $H_{\rm c2}$.
This enhanced magnetostriction above $H_{\rm c2}$ is consistent with our theoretical calculation, as shown in Fig.~2(d) of the main text.

\vspace{+1.0cm}
\subsection*{\label{Sec6} \large Note 6. Temperature dependence of calculated magnetization and magnetostriction curves}

\begin{figure}[h]
\centering
\vspace{+0.5cm}
\includegraphics[width=0.8\linewidth]{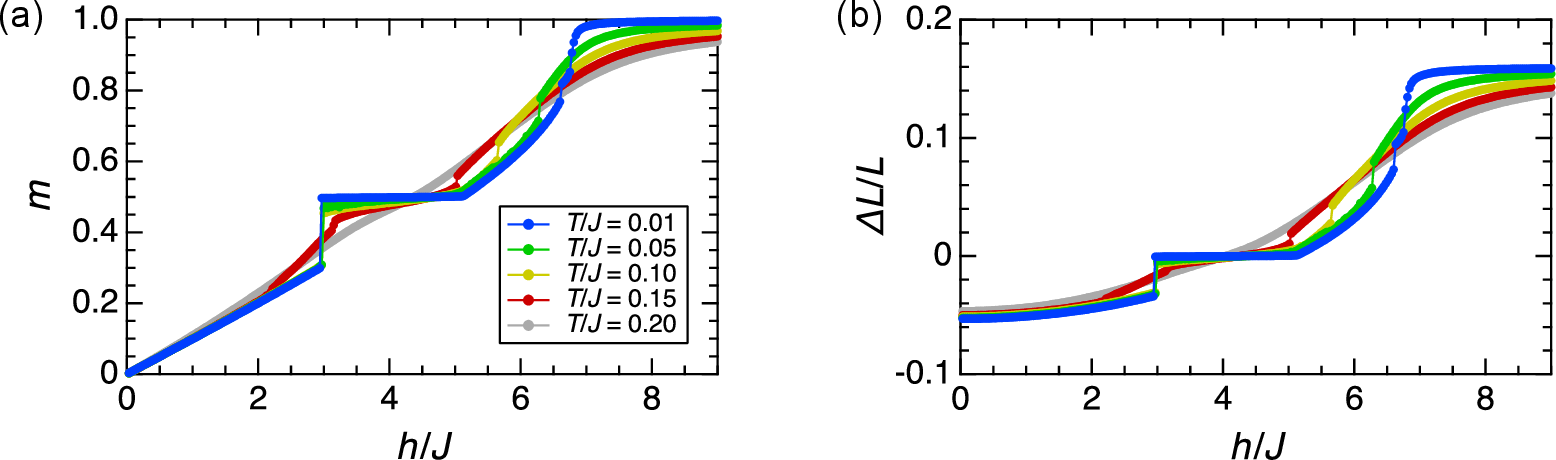}
\caption{Calculated (a) magnetization and (b) magnetostriction curves for the extended SLC model [Eq.~\eqref{EqS1}] with $b = 0.2$ and $\eta = 0.6$ at various temperatures.}
\label{FigS4}
\end{figure}

Figure~\ref{FigS4} shows the calculated magnetization and magnetostriction curves with $b = 0.2$ and $\eta = 0.6$ at various temperatures.
The double-peak structure in $dM/dH$ just below saturation is relatively broadened for CdCr$_{2}$O$_{4}$ ($T_{\rm N} = 7.8$~K) at $T_{\rm ini} = 4.2$~K [Fig.~2(a)], compared to the calculated result at $T/J = 0.01$ [Fig.~2(c)].
This difference can be attributed to thermal fluctuations, given that in our calculations the two-step metamagnetic anomaly already disappears upon increasing the temperature above $T/J = 0.05$.
In addition, Fig.~\ref{FigS4}(a) shows that the sign of ${\partial M}/{\partial T}$ changes from negative to positive around $h/J = 4.5$, which is higher than the midpoint of the 1/2-plateau phase, $h/J \approx 4.1$.
This behavior is consistent with the trend of the distorted dome-like structure of the MCE curves (see Fig.~4(a) and the related discussion in the main text for details).

\clearpage
\subsection*{\label{Sec7} \large Note 7. Magnetic structure of the HF phase}

\begin{figure}[h]
\centering
\vspace{+1.0cm}
\includegraphics[width=0.8\linewidth]{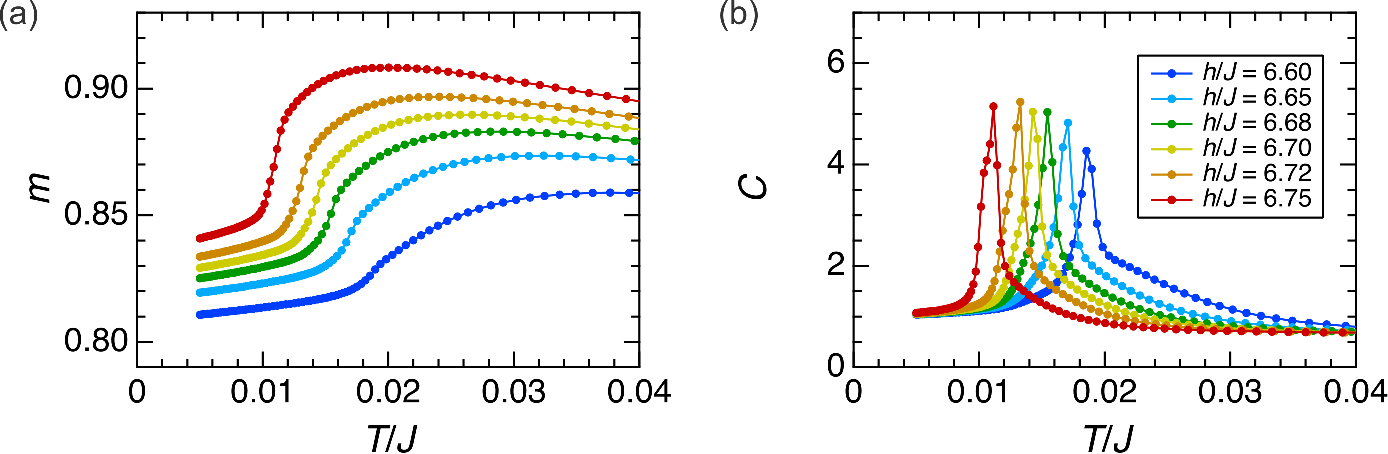}
\caption{Temperature dependence of (a) magnetization and (b) specific heat for the extended SLC model [Eq.~\eqref{EqS1}] with $b = 0.2$ and $\eta = 0.6$ at various magnetic fields in the HF phase.}
\vspace{+0.7cm}
\label{FigS5}
\end{figure}

Here, we focus on the HF phase, which is absent in the pure bond-phonon or site-phonon models, but appears in the extended SLC model [Eq.~\eqref{EqS1}], as shown in the phase diagram in Fig.~2(e) of the main text.
The HF phase emerges between the canted 3:1 phase and the fully-polarized phase.
Figure~\ref{FigS5} shows the calculated temperature dependence of magnetization and specific heat at high magnetic fields in the range $h/J = 6.60$--6.75 with $b = 0.2$ and $\eta = 0.6$. 
A clear signature of a magnetic transition from the paramagnetic phase to a magnetically ordered phase is indicated by a sharp peak in the specific heat.
The transition temperature gradually decreases with increasing magnetic field, reaching $T_{\rm N}/J \approx 0.01$ at $h = 6.75$.

\begin{figure}[t]
\centering
\includegraphics[width=0.98\linewidth]{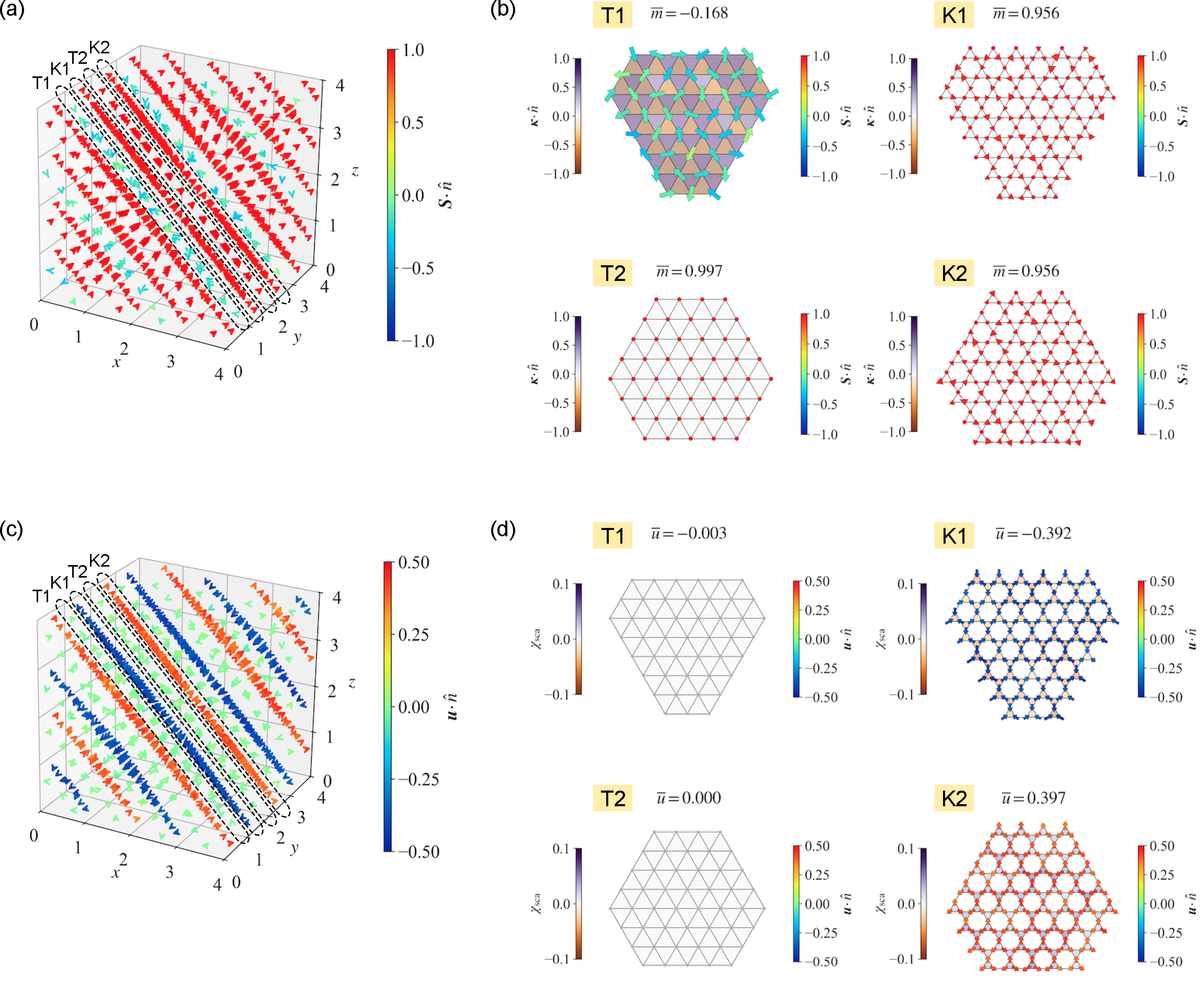}
\caption{MC snapshots of the [(a)(b)] spin configurations and [(c)(d)] lattice-distortion patterns obtained at $h/J = 6.65$ and $T/J = 0.005$ for the extended SLC model [Eq.~\eqref{EqS1}] with $b = 0.2$ and $\eta = 0.6$. For clarity, the magnetic field is applied along the $\langle 111 \rangle$ direction instead of the $z$ axis. Panels (b) and (d) show four consecutive triangular (T1 and T2) and kagome (K1 and K2) layers perpendicular to the $\langle 111 \rangle$ axis, as indicated in panels (a) and (c). In panels (a) [(c)], the arrow directions indicate the spins (lattice displacements) in the pyrochlore lattice; in panels (b) [(d)], the arrow directions indicate the in-plane components, and the arrow colors represent the out-of-plane components, $\boldsymbol{S} \cdot \hat{n}$ ($\boldsymbol{u} \cdot \hat{n}$), where $\hat{n}$ is the unit vector in the $\langle 111 \rangle$ axis. In panel (b) [(d)], the out-of-plane component of the vector spin chirality, $\boldsymbol{\kappa} \cdot \hat{n}$, (the scalar chirality of lattice displacements, $\chi_{\rm sca}$) is visualized by color for each triangle. The layer-resolved averages of the spin and lattice-displacement components parallel to the field direction are also displayed as $\bar{m}$ and $\bar{u}$, respectively.}
\label{FigS6}
\end{figure}

Figures~\ref{FigS6}(a) and \ref{FigS6}(c) show MC snapshots of the spin configurations and lattice distortions, respectively, in the HF phase.
Figures~\ref{FigS6}(b) and \ref{FigS6}(d) present the corresponding layer-resolved plots for four consecutive layers: two triangular layers (T1 and T2) and two kagome (K1 and K2) layers alternately stacked along the $\langle 111 \rangle$ direction.
These results reveal the emergence of a magnetic LRO state characterized by a periodic stacking sequence of a triangular layer with a 120$^{\circ}$ spin configuration (T1), an all-up kagome layer (K1), an all-up triangular layer (T2), and another all-up kagome layer (K2).
Here, the out-of-plane component of the spins in the T1 layer takes a negative value, oriented opposite to the magnetic field direction.
Interestingly, in the T1 layer, the in-plane components of the spins form a 120$^\circ$ configuration, exhibiting an antiferrochiral order in which neighboring T1 layer possess opposite vector spin chirality.
On the site-phonon side, breathing-type in-plane lattice displacements appear in the kagome layers, while both adjacent kagome layers are displaced toward the T1 layer along the out-of-plane direction.

\newpage

\clearpage
\subsection*{\label{Sec8} \large Note 8. Specific heat data for polycrystalline CdCr$_{2}$O$_{4}$ in high magnetic fields}

\begin{figure}[h]
\centering
\vspace{+0.5cm}
\includegraphics[width=0.45\linewidth]{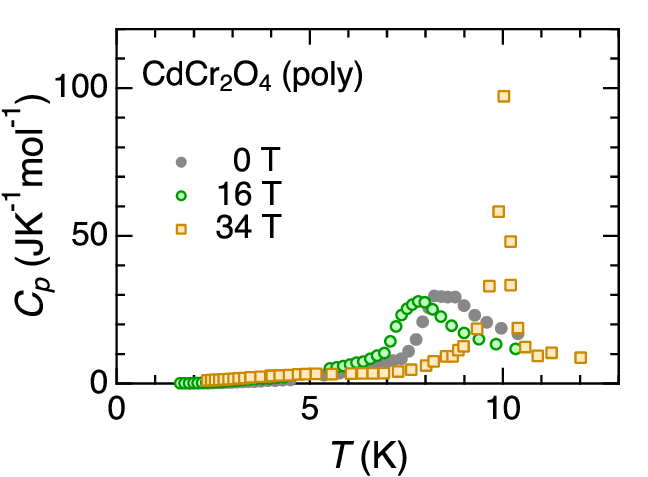}
\caption{Temperature dependence of specific heat at 16~T and 34~T for polycrystalline CdCr$_{2}$O$_{4}$.}
\label{FigS7}
\end{figure}

Figure~\ref{FigS7} shows the specific heat data measured on polycrystalline sintered samples of CdCr$_{2}$O$_{4}$ at 16~T and 34~T, obtained using a flat-top long pulsed magnet \cite{2015_Koh, 2021_Ima}.
Compared to the single-crystal data shown in Fig.~4(c) of the main text, the specific heat peak associated with the phase transition is notably broadened, particularly at lower fields.
However, the peak corresponding to the transition to the 3-up--1-down LRO phase at 34~T is remarkably sharp, as observed on HgCr$_{2}$O$_{4}$ \cite{2022_Kim}.

\vspace{+1.0cm}
\subsection*{\label{Sec9} \large Note 9. Thermodynamic properties on the high-field side of the 1/2-plateau phase}

\begin{figure}[h]
\centering
\vspace{+0.5cm}
\includegraphics[width=0.95\linewidth]{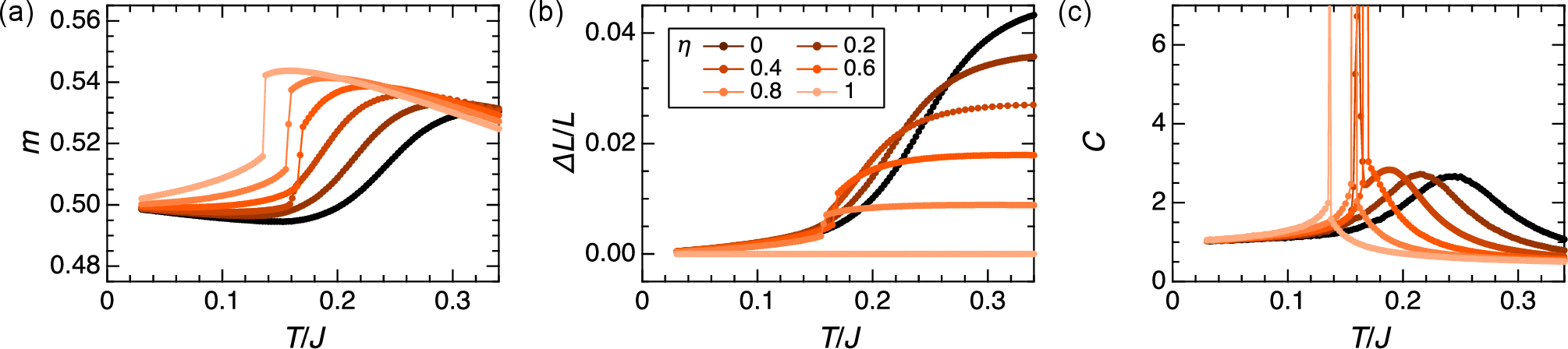}
\caption{Temperature dependence of (a) magnetization, (b) thermal expansion, and (c) specific heat for the extended SLC model [Eq.~\eqref{EqS1}] with $b = 0.2$ and various $\eta$ values at $h/J = 4.7$.}
\vspace{+2.0cm}
\label{FigS8}
\end{figure}

Figure~\ref{FigS8} shows the calculated temperature dependence of magnetization, thermal expansion, and specific heat for the extended SLC model [Eq.~\eqref{EqS1}] with $b = 0.2$ and various $\eta$ values in the high-field side of the 1/2-magnetization plateau at $h/J = 4.7$.
In contrast to the low-field side of the 1/2-plateau, the thermal expansion exhibits a positive temperature dependence.

